\documentclass[journal=jacsat,manuscript=article]{achemso}
\usepackage{hyperref}
\usepackage{physics,amsmath}
\usepackage{soul,color}
\SectionNumbersOn

\usepackage{ulem}
\usepackage[version=3]{mhchem} 

\author{Arpan Choudhury}
\affiliation{School of Chemical Sciences, Indian Association for the Cultivation of Science, \\ Kolkata, India}
\altaffiliation{These authors contributed equally.}
\author{Sonaldeep Halder}
\affiliation{Department of Chemistry, Indian Institute of Technology Bombay, \\ Mumbai, India}
\altaffiliation{These authors contributed equally.}
\author{Rahul Maitra}
\affiliation{Department of Chemistry, Indian Institute of Technology Bombay, \\ Mumbai, India}
\email{rmaitra@chem.iitb.ac.in}
\author{Debashree Ghosh}
\affiliation{School of Chemical Sciences, Indian Association for the Cultivation of Science, \\ Kolkata, India}
\email{pcdg@iacs.res.in}

\title[An \textsf{achemso} demo]
  {
  Fragment, Entangle, and Consolidate: Strong Correlation through Bi-fold Quantum Circuits}

\abbreviations{IR,NMR,UV}

\begin{document}







\begin{abstract}
An accurate description of strong correlation is quintessential for the exploration of emerging chemical phenomena. While near-term variational quantum algorithms provide a theoretically scalable framework for quantum chemical problems, the accurate simulation of multireference effects remains elusive, hindering progress toward the rational design of novel chemical space. In this regard, we introduce a general and customizable scheme to handle strong electronic correlation, based on problem decomposition, entanglement buildup, and subsequent consolidation. Based on a problem-inspired molecular decomposition, the deployment of Hardware Efficient Ansatz to prepare entangled subsystems ensures efficient construction of a multireference state while concurrently adhering to the hardware topology. The dynamic correlation is subsequently introduced through a unitary coupled cluster framework, with static or dynamic ansatz parametrized by a set of inter-fragment generalized operators, and with the product state spanning various subsystems taken as the reference. The hybrid architecture ensures a judicious deployment of separate ansatze structures for capturing various degrees of correlation in a balanced manner, while concurrently retaining the scalability and flexibility provided by them individually. Over a number of numerical applications on a strongly correlated system, the proposed scheme is shown to be highly accurate, flexible, and robust in unlocking the potential to harness quantum advantage for quantum chemistry.   
\end{abstract}

\section{Introduction}
The exact, or full configuration interaction (FCI), solution of the electronic Schr\"odinger equation exhibits an exponential scaling with system size. Consequently, a variety of approximate approaches—such as M\o ller–Plesset perturbation theory, truncated configuration interaction, and coupled-cluster theory—have been developed to tackle electronic structure problems.
Even in these approximate algorithms, the computational complexity grows rapidly with the system's size, which makes their manipulation using classical computers intractable.\cite{simons2023quantum}
Quantum computing platforms promise a 
quantum advantage for eigenstate calculations\cite{abrams1999quantum,lee2023evaluating}, which can have
tremendous consequences in the fields of drug discovery, novel
material design, and synthesis of efficient catalysts. The 
currently available pre-fault-tolerant quantum devices are limited
in their number of qubits, coherence times, and suffer from various
sources of noise. Under these constraints, the Variational Quantum 
Algorithm (VQE)\cite{peruzzo2014variational,bharti2022noisy,delgado2021variational, 
grimsley2019adaptive,halder2022dual,
halder2023corrections, 
mondal2023development,feniou2023overlap,zhao2023orbital, tang2021qubit, yordanov2021qubit, ostaszewski2021structure, tkachenko2021correlation, zhang2021adaptive, sim2021adaptive} is one of the leading
methodologies for simulating molecular systems on current quantum 
hardware. It involves a parametric construction  
of a wavefunction on a quantum hardware and subsequent classical optimization. However, as the theoretical foundation has evolved, a number of its drawbacks have surfaced. This includes the contradictory issues of trainability and expressibility\cite{sim2019expressibility, truger2024warm, tilly2022variational, jattana2023improved}; the appearance of barren plateaus\cite{larocca2024review, mcclean2018barren, anschuetz2022quantum, qi2023barren, arrasmith2022equivalence, larocca2022diagnosing, cerezo2023does,cybulski2023impact, ragone2023unified, fontana2023adjoint, bremner2009random, gross2009most} where the optimization landscape becomes exponentially flat with the system size prohibits us from making an ansatz arbitrarily expressible. Even without considering the hardware limitations, the very theoretical construct of VQE makes it difficult to be applicable for realistic molecular systems where strong correlations plays an important role due to their inherent limitations to reach the ground state when the full circuit is expressed over all the qubits. 
Thus, designing a shallow depth ansatz adhering to the hardware constraints which efficiently captures strong correlation effects is at the heart of modern quantum chemistry research. 
Through this work, we aim to overcome some of these theoretical bottlenecks 
through a many-body approach while aiming for an
accurate determination of molecular energetics for moderate to
strongly correlated systems.

The core of our approach lies in, first, forming a product state
composing of individual entangled states. These individual states
could be formed using a shallow depth ansatz, such as the Hardware
Efficient Ansatz (HEA) acting on sub-parts of the whole system. The 
classification into these sub-parts could be pivoted upon a suitable problem decomposition approach, like the construction of disconnected subsystems based on orbital symmetries or orbital localizations.
The HEA is applied only over sub-parts with substantially lower variational parameters as compared to full system HEA. This is 
targeted to scale down the barren plateau problem suffered by 
HEA as the system size increases. Moreover, when integrated as small 
building blocks within more sophisticated circuit architectures—
such as quantum circuit tensor networks— HEA can achieve ground 
state energies within chemical accuracy (i.e. $\le 1$
kcal/mol)\cite{liu2019variational,haghshenas2022variational,fan2023quantum}, bringing in additional designing flexibility. 
The individual sub-parts can be variationally optimized 
in parallel. From the correlation point of view, the HEA generates
a strongly correlated initial state, spanning over the entire 
Hilbert space as a tensor product of individual entangled 
sub-Hilbert spaces.

The second step involves building the inter-fragment 
entanglement between these sub-parts using a judiciously 
chosen ansatz. The dUCC \cite{evangelista2019exact} ansatz with customized 
inter-fragment excitation operators may be the optimal choice for 
this purpose, as such a representation is known to accurately 
capture the dynamic correlation. 
The purposes of this two-step approach bring in great flexibility in 
ansatz design: in particular, chemistry-inspired fragmentation allows us to 
handle the subsystems separately in a restricted qubit space with any ansatz 
of choice adhering to the hardware coupling map, and thus to span the full 
Hilbert space through the multireference product state (MRPS).
The use of this MRPS, as will be seen in the later
sections, becomes imperative to accurately describe chemical
phenomena such as bond breaking. In such scenarios, single
reference dUCC-based ansatze often prove to be insufficient. The subsystem-wise
optimization also brings in additional flexibility to include dynamic
correlation only through inter-fragment (generalized) operators, and thus reduces the parameter count tremendously. Moreover, the fragmentation 
also allows us to include static and dynamic correlation through two 
different ansatz in two disconnected optimization cycles, although one may 
optionally relax the MRPS after the inclusion of dynamic correlation in an 
iterative manner. Such a disconnected optimization protocol through two
disjoint cycles reduces the exponential complexity of optimization for 
large chemical systems within practical realization. We henceforth use the term bi-folded approach which refers to our two-step computational strategy for modeling electron correlation. The first `fold' involves dividing the system into smaller fragments and preparing a correlated product state from them. The second `fold' then systematically incorporates the inter-fragment correlations to describe the complete molecular system. This separation of intra-fragment and inter-fragment effects provides an efficient and scalable pathway for simulating large, strongly correlated molecules.


One of the earliest approaches in this direction, involves using adiabatic state preparation.\cite{aspuru2005simulated,du2010nmr,veis2014adiabatic}
Later on, Gagliardi and co-workers introduced an approach where, instead 
of starting from a single configurational Hartree-Fock (HF) reference, the 
initial state is prepared using a cluster mean-field method known as the 
localized active space self-consistent field. This leverages molecular 
substructures or fragments. \cite{hermes2019multiconfigurational,otten2022localized,d2024state}
The individual fragments are solved using quantum phase estimation, 
while the correlations between fragments are incorporated using the UCC.
While this is conceptually elegant and theoretically precise, the use of a phase estimation algorithm necessitates the availability fault-tolerant device.
Another notable development in the multi-determinantal VQE was put forward 
by Sugisaki and co-workers, where the multiconfigurational state was prepared 
on classical computers without performing any post-HF calculations.\cite{sugisaki2018quantum} While the UCC-based disentangled static ansatz is conceptually simple, it has its limitations, in particular, its heuristic nature dictated by the ordering of the operators. Machine learning techniques 
have recently been employed to assist in designing dUCC-based ans\"atze for 
electronic structure simulations\cite{halder2024machine,halder2025construction,patel2024curriculum}, 
particularly in numerically deciphering a correct operator ordering for improved
energy accuracy. Contrarily, a dynamically
designed ansatz based on finite order pseudo-Trotter expansion\cite{grimsley2019adaptive}
leverages the brute-force optimization of the cost function numerically at the cost of
exponentially high pre-circuit measurements. 
Moreover, methods based on qubit excitations, like qubit UCC and qubit-ADAPT-VQE, 
are particularly appealing when the objective is to build shallower circuits 
\cite{ryabinkin2018qubit,tang2021qubit} and have been proved to be successful in the
context of NISQ computing. However, none of these approaches
had been exhaustively tested on multireference initial states, which potentially
may overcome their inaccuracy in the regions of strong correlation-- a caveat 
posed due to a poor optimization landscape and poor parametrization. In this work,
we put forward a major step to overcome these limitations by breaking down the overall
problem into smaller sub-problems and build-up VQE, with both static and dynamically 
constructed ansatze, with a multireference initial (product) state to handle electronic 
strong correlation. We demonstrate the superiority of this hybrid bi-folded scheme
with several applications on strongly correlated systems, including the 
quantum-enabled calculation of the reaction barrier for cyclobutadiene automerization.

\section{Methods}

There exists a significant number of important chemical phenomena that exhibit multireference character. This situation typically occurs when there are nearly degenerate orbitals, resulting in inadequate description through a single Slater Determinant reference. Some typical examples include understanding of electronic excited states, transition metal complexes, bond breaking, and systems with unpaired electrons. In the quantum computing platform, the development of such multireference methods, which can also navigate the bottlenecks put forth by the hardware, is a challenging task. These bottlenecks include the presence of noise, limited coherence times, barren-plateau problems, and the requirement of shallow circuits, which make best use of the underlying topology of the hardware. A very lucrative approach is to break down the molecular system into subparts and employ a bi-fold correlation capture scheme through the use of efficient and customizable ansatz structures. This division would allow one to utilize separate circuit structures -- one to bring entanglement within each sub-part, and another to capture correlations between these sub-parts. The HEA suffers from barren plateau problems where the gradients of the cost function vanish exponentially with the number of qubits. The dUCC-based ansatze result in relatively deep circuits. Moreover, a single ansatz may not be efficient for multireference systems. This work uses physically motivated partition techniques enabled by orbital localization or orbital symmetry to build a
\textit{multireference product state}. 
This utilizes the strengths of HEA while mitigating the barren plateau problem due to its implementation across subparts rather than the entire system. This multireference state is further entangled through an adaptive ansatz, which only searches for correlations between different subparts, thereby reducing the cost required for searching for correlations throughout the full system. 
In this section, we begin by introducing the construction of an MRPS where entanglement is introduced only within individual subparts, after partitioning the system. This is followed by a description of the 
incorporation of inter-subpart correlation using a shallow adaptive ansatz. This leads to the Results and Discussion section, where we numerically analyse the efficacy and cost of our technique.

\subsection{Generation of the Multireference Product State}

The dimension of the complete active space that captures the 
multireference effects over the entire molecule grows exponentially with the system size. This approach aims to scale down the 
exponential barrier by decomposing the system into several chemically motivated sub-parts to entangle orbitals within individual sub-systems. Such problem decomposition also ensures that each individual sub-system is efficiently optimized as they span only a smaller sector of the Hilbert space over a restricted number of qubits. 
This results in improved tractability of the unitary ansatz used to describe correlations within individual subsystems, while also allowing each subsystem to be prepared independently and simultaneously on parallel quantum architectures, provided a sufficient number of quantum devices are available.


For chemistry-inspired problem decomposition, spatial localization or orbital symmetries can be used as they optimally capture system-specific chemical features. We briefly discuss our scheme towards the fragmentation of the system:

Starting with an HF calculation on the entire system, two different approaches are employed to decompose the system into smaller subsystems. In the first approach, based on orbital symmetry, canonical orbitals corresponding to distinct symmetries (arising from the molecular point group) are assigned to different subsystems.
In the second approach, a set of localized molecular orbitals (LMOs) is obtained by expressing them as linear combinations of canonical molecular orbitals (CMOs) as
\begin{gather}
    \widetilde{\phi}_{m} (\mathbf{r}) = \sum_n \phi_n (\mathbf{r}) U_{nm}
\end{gather}
where $U_{nm}$ form a unitary rotation. 
Then, the subsystems are defined spatially based on the LMOs.
As there is ambiguity in the choice of localization procedure, several approaches to constructing LMOs from CMOs have been proposed, among which the Pipek-Mezey,\cite{pipek1989fast} Boys,\cite{foster1960canonical}, and natural atomic orbitals\cite{reed1985natural} methods are common. The problem decomposition allows us to map different subsystems parallelly onto the qubits.

If the system is divided into $N_{subsys}$ subsystems, 
the entangled states for each subsystem are prepared using an embedding technique. Here we used the embedding proposed in Ref. \citenum{rossmannek2021quantum}, where the effective Hamiltonian ($\hat{H}$) for one subsystem (say $\mathbf{A}$) becomes:
\begin{gather}
    \hat{H}^\mathbf{A} = \sum _{uv \in \mathbf{A}} F^\mathbf{A}_{uv} \hat{a}^\dagger_{u} \hat{a}_{v} + \frac{1}{2} \sum_{uvxy \in \mathbf{A}} g_{uvxy} \hat{a}^\dagger_{u} \hat{a}^\dagger_{v} \hat{a}_{x}\hat{a}_{y}
    \label{eq:eff-ham}
\end{gather}
and the \textit{Fock operator} $F^\mathbf{A}_{uv}$ is defined as
\begin{gather}
    F^\mathbf{A}_{uv} = h_{uv} + \frac{1}{2} \sum_{i \in \mathbf{A}\bigoplus\mathbf{B}} \Big( 2g_{iiuv} -g_{ivui}\Big).
    \label{eq:fock}
\end{gather}
Here, $\mathbf{A}$ represents the subsystem of interest and $\mathbf{B}$ represents the collection of the remaining subsystems. 
$h_{pq}$ and $g_{pqrs}$ are the one- and two-electron integrals, respectively.


The optimized entangled state of subsystem $\mathbf{A}$, which we write as $|\psi^{\mathbf{A}}(\vec{\theta}) \rangle \in \mathcal{H}^A$ ($\mathcal{H}^A$ is the Hilbert space of subsystem $\textbf{A}$) is prepared by minimizing the corresponding energy expectation value with the embedding Hamiltonian:
\begin{gather}
    E^\mathbf{A} (\vec{\theta}^A)= \langle \psi^{\mathbf{A}}(\vec{\theta}^A)|\hat{H}^\mathbf{A}|\psi^{\mathbf{A}}(\vec{\theta}^A) \rangle.
    \label{eq:hea-energy}
\end{gather}
A similar procedure can be carried out in parallel for the remaining subsystems to prepare their entangled states.

\begin{figure}[!htb]
    \centering    \includegraphics[width=\linewidth]{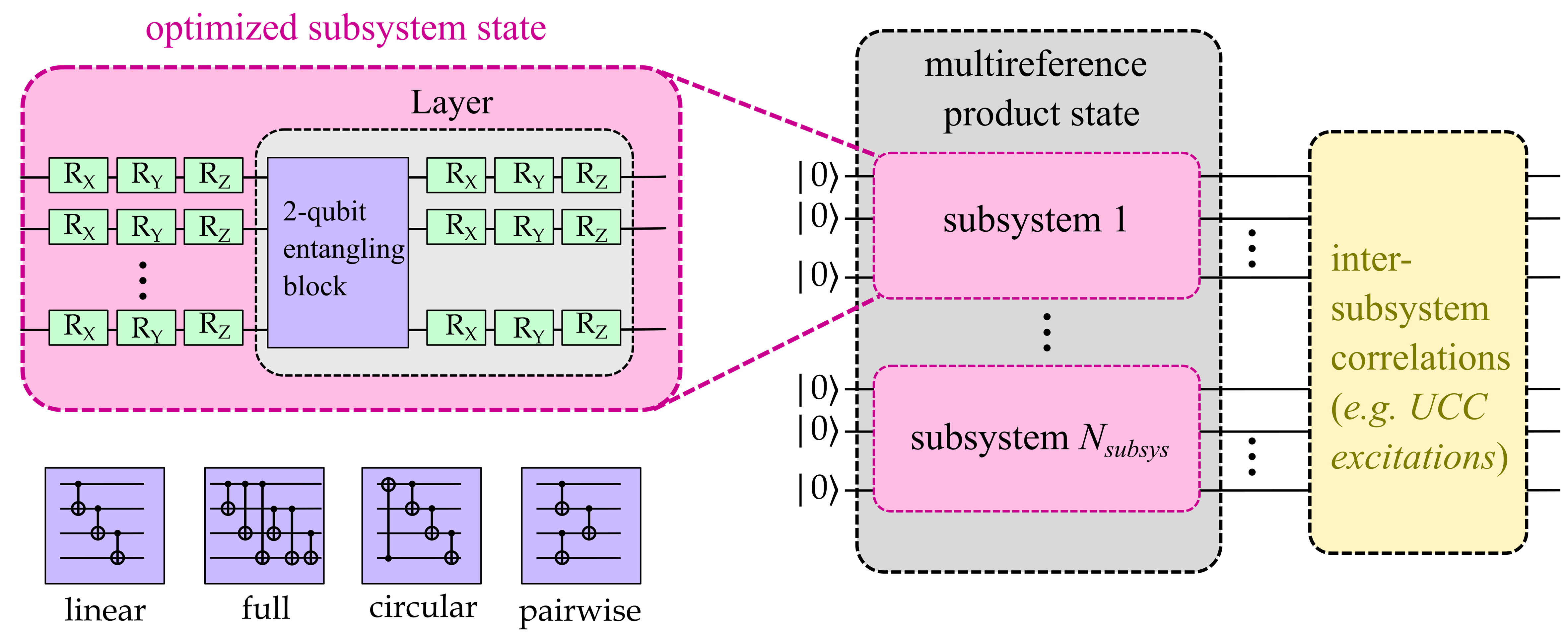}
    \caption{The procedure begins by constructing a multireference product state through parallel optimization of the subsystems, for instance, using HEA (as shown on the left side). Subsequently, inter-subsystem correlations are incorporated to recover the complete wavefunction with methods such as ADAPT-VQE.}
    \label{fig:hea-qubit-scheme}
\end{figure}

For each $N_A$-qubit ($N_A << n$) subsystem, the entangled state can be prepared using an ansatz that is attuned to the hardware topology, such as the HEA ansatz:
\begin{gather} 
    |\psi^\mathbf{A}(\vec{\theta}) \rangle = \prod_{l=1} ^{L_A} U_l^\mathbf{A}(\theta _l^\mathbf{A}) \ |\phi_0^\mathbf{A} \rangle 
\end{gather}
where $|\phi_0^\mathbf{A} \rangle \in \mathcal{H}^A$ is a reference state over the subsystem $\mathbf{A}$ (which can be taken as all the $N_A$ qubits initialized to $|0\rangle$) and $U_l(\vec{\theta _l^\mathbf{A}})$ is a unitary, acting only on states of $\mathcal{H}^A$, consisting of gates which are native to quantum hardware, such as single-qubit parameterized gates and two-qubit entangling gates (see Fig. \ref{fig:hea-qubit-scheme}). Furthermore, to increase the variational flexibility in capturing the intra-subsystem correlation, the circuit may have $L_A$ repeating layers of $U_l(\vec{\theta _l^\textbf{A}})$. 
The tensor-product state at this stage, denoted as the MRPS ($|\psi_\mathrm{MRPS}(\vec{\theta}) \rangle \in \bigotimes_{\mathbf{K} =1}^{N_{subsys}}\mathcal{H}^\mathbf{K}$), which resides in the Hilbert space of the entire molecule, can be written as:
\begin{align} \label{product_state}
|\psi_\mathrm{MRPS}(\vec{\theta}) \rangle &= \bigotimes_{\mathbf{K} =1}^{N_{subsys}}|\psi^\mathbf{K}(\vec{\theta}^{\mathbf{K}}) \rangle \\
&= \bigotimes_{\mathbf{K} =1}^{N_{subsys}}\prod_{l=1} ^{L_K} U_l^\mathbf{K}(\theta _l^\mathbf{K}) \ |\phi_0^\mathbf{K} \rangle 
\end{align}
Decomposing the problem into multiple embedded subsystems through subsystem-specific unitaries provides significant flexibility in designing the circuit structure of $U_l^K(\vec{\theta^K_l})$ within the hardware constraints. In our case, the single-qubit rotation gates $R_X$, $R_Y$, and $R_Z$ are used as parameterized gates, and the two-qubit controlled Pauli gates, such as CNOT, are used as entangling gates. Various structures of the entangling layers are explored, and they are shown in Fig. \ref{fig:hea-qubit-scheme}. 

\subsection{Incorporation of Inter-Subsystem Correlations}
While the MRPS captures a strong correlation with the embedded Hamiltonian over disjoint fragments, it is 
imperative that one entangles the different fragments without further
relaxing the quantum states over individual subsystems. Thus with the
optimized MRPS taken as the reference, we strategically incorporate the inter-subsystem 
correlation.
We choose to do so by utilizing the Adaptive Derivative-Assembled
Pseudo-Trotter ansatz Variational Quantum Eigensolver (ADAPT-VQE) scheme \cite{grimsley2019adaptive}. 
The choice of the operator pool plays a crucial role in ADAPT-VQE and 
such a choice needs to be explored for MRPS reference states. 
Let $|\psi_\mathrm{MRPS}(\vec{\theta}_{opt}) \rangle$ be an optimized product state (Eq. \eqref{product_state}) where each entanglement block is optimized via Eq. \eqref{eq:hea-energy}. Performing ADAPT-VQE, taking $|\psi_\mathrm{MRPS}(\vec{\theta}_{opt}) \rangle$ as the reference state; after $m$ iterations, the state $|\psi(\vec{\theta}) \rangle \in \bigotimes_{\mathbf{K} =1}^{N_{subsys}}\mathcal{H}^\mathbf{K}$ ($\mathcal{H}^\mathbf{I}$ is the Hilbert space associated with $I^{th}$ subsystem)  can be written as:

\begin{gather}
    |\psi(\vec{\theta}) \rangle = e^{\theta_m \hat{\tau}_m} \dots e^{\theta_i \hat{\tau}_i} \dots e^{\theta_2 \hat{\tau}_2} e^{\theta_1 \hat{\tau}_1} |\psi_\mathrm{MRPS}(\vec{\theta}_{opt}) \rangle \\
    = \prod_{k=1}^m e^{\theta_k \hat{\tau}_k} |\psi_\mathrm{MRPS}(\vec{\theta}_{opt}) \rangle
    \label{eq:adapt}
\end{gather}
where $\hat{\tau}_k$ are fermionic single or double anti-hermitian ``excitation" operators defined as:
\begin{eqnarray}
    \tau_p^{q}= \hat{a}_q^{\dagger} \hat{a}_p - \hat{a}_p ^\dagger\hat{a}_q \\
    \tau_{pq}^{rs}= \hat{a}_s ^\dagger \hat{a}_r ^\dagger \hat{a}_p \hat{a}_q - \hat{a}_q ^\dagger \hat{a}_p ^\dagger \hat{a}_r \hat{a}_s.
\end{eqnarray}
Here, $p,q,r,s$ are generalized orbital indices. Importantly, the generalized operators
$\{\tau_I\}_I$ act only across subsystems, that is, at least one of the indices in both single and double ``excitation" operators must belong to another subsystem compared to the rest. 
Since ADAPT-VQE scans the entire operator pool for selecting the optimum ones, a reduction in the size of the operator pool translates directly to a lowering of the required measurements.
It is important to note that in Eq. \eqref{eq:adapt}, the variational parameters are the ones present in the exponential operators. While performing the ADAPT-VQE, the full molecular Hamiltonian is used both in determining the gradients for selecting the operators $\tau_i$ and for parameter optimization associated with the exponential ansatz.
One can utilize Qubit-ADAPT-VQE, a variant of conventional ADAPT-VQE, to bring in the inter-subsystem correlations. A detailed description of Qubit-ADAPT-VQE is given in Supplementary Information.

In particular, combining the HEA ansatz on individual subsystems followed by the use ADAPT-VQE to bring inter-subsystem correlations provides a complementary approach that overcomes the limitations of using either method alone. The HEA suffers from barren plateau problem where the loss function (the energy function) becomes exponentially concentrated with the system size. For an $n$ qubit system, the variance of the cost function $E(\boldsymbol{\theta})$ scales as\cite{mcclean2018barren}:

\begin{gather}
\text{Var}_{\boldsymbol{\theta}}\left(\frac{\partial E (\boldsymbol{\theta})}{\partial \theta_k}\right) \in O\left(\frac{1}{b^n}\right), \space b >1
\end{gather}

where $\theta_k \in \boldsymbol{\theta}$. For an HEA acting on $N_A$ qubit subsystem A, the variance scales
as:

\begin{gather}
    O(\frac{1}{b^{N_A}}), \space N_A<<n
\end{gather}
This, in practice, alleviates the bottleneck present for the HEA ansatz within the molecular Hamiltonian. 

The following Results and Discussion section is structured to first systematically validate the individual components of our bi-folded approach, thereby establishing its foundational efficacy. With this groundwork laid, we then demonstrate the full power of the integrated methodology through its application to a series of challenging, strongly correlated molecular systems.

\section{Results and Discussion}

\begin{figure}[!htb]
    \centering    \includegraphics[width=\linewidth]{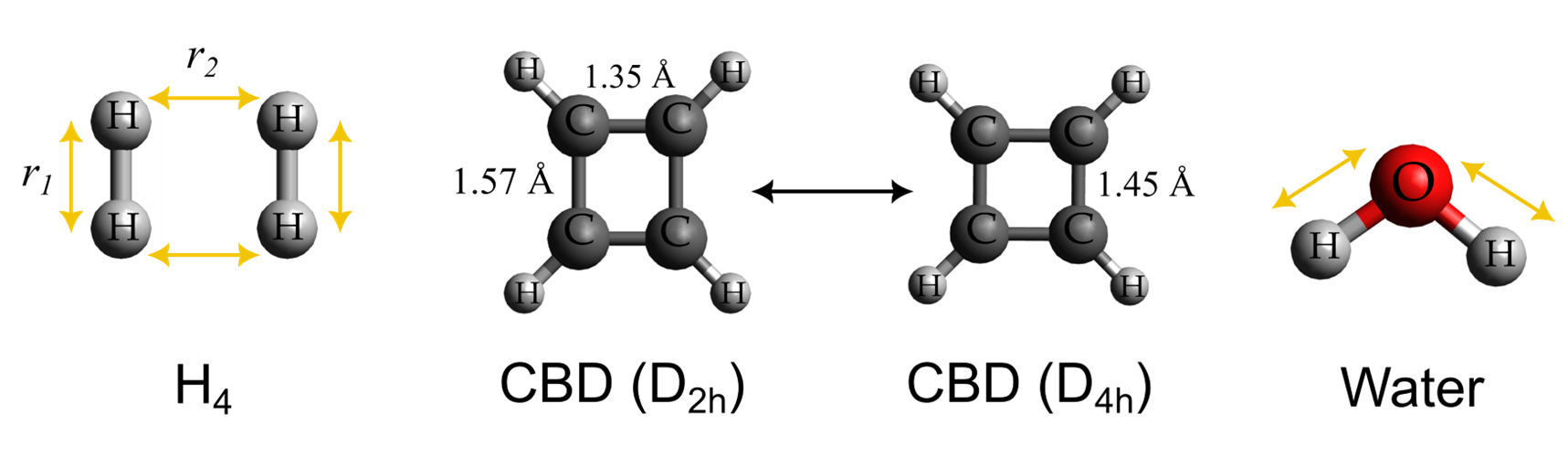}
    \caption{Models and molecular systems investigated in this work: H$_4$, cyclobutadiene (CBD) in $D_{2h}$ and $D_{4h}$ symmetries taken from Ref.\citenum{monino2022reference}, and water.}
    \label{fig:molecule}
\end{figure}

We first analyze the efficacy of initial state preparation by VQE optimizations. 
Then, the performance of full UCCGSD, fermionic-ADAPT-VQE and qubit-ADAPT-VQE on an initial MRPS state, as well as a HF state, is discussed. 
The results are shown for various molecular systems as depicted in Fig. \ref{fig:molecule}. 
We used Molpro\cite{werner2012molpro} for performing orbital localization and classical multireference calculations, and PySCF\cite{sun2018pyscf} for the one- and two-electron integrals computation. 
For all VQE calculations, we used Qiskit simulators \cite{Qiskit} with Limited-memory Broyden-Fletcher-Goldfarb-Shanno Bound (L-BFGS-B) optimizer. 
The STO-3G basis set is used for H$_4$ systems and water, and the 6-31G basis set is used for cyclobutadiene (CBD).
The exact results (i.e., FCI within the chosen basis set) are obtained by performing exact diagonalization of the qubit Hamiltonian.

\subsection{Convergence of the Fragment States}

The fragment states are prepared using the effective Hamiltonians (see Eq. \ref{eq:eff-ham}), where each fragment’s Hamiltonian contains an embedded Coulomb potential due to the other fragments. 
Consequently, the fragments are not independent, and the sum of their individual energies does not correspond to the total energy of the full system.
For a simple H$_4$ system in both rectangular ($r_1 =1.0  $\AA, $r_2 = 2.0$\AA) and square ($r_1  =r_2 = 1.0$\AA) geometries, two fragment states are prepared separately on two H$_2$ moieties. Each H$_2$ fragment state corresponds to a 4-qubit system, which is prepared using LMOs. 
For the CBD molecule, we used a (4o,4e) active space in the 6-31G basis set, which includes only the $\pi$ and $\pi^*$ MOs. 
These CMOs are localized using the Pipek-Mezey scheme to obtain the LMOs.
The individual fragment states on each C$_2$H$_2$ are prepared using the LMOs in a (2o,2e) fragment active space.

\begin{figure}[!htb]
    \centering
        \includegraphics[width=\linewidth]{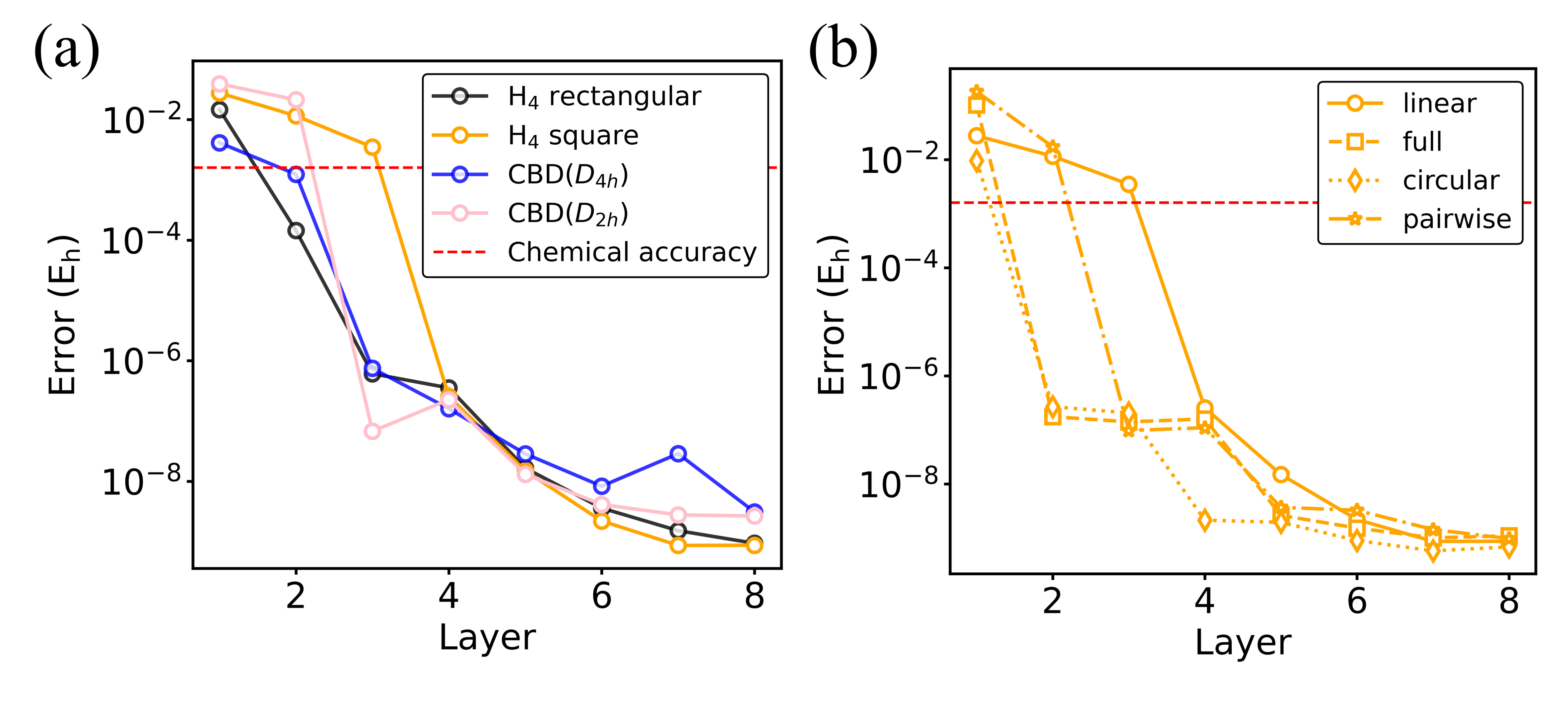}
    \caption{(a) The error (in Hartree) in fragments’ state energy with respect to the exact energy (obtained via exact diagonalization of the corresponding fragment Hamiltonian) against the number of layers in HEA using linear entanglement blocks.
 (b) The same error for the square H$_4$ geometry with the number of layers using linear, full, circular, and pairwise entanglement blocks. It is to be noted that the Fock operator of the individual fragment Hamiltonian has an embedded Coulomb potential of the other fragments (Eq. \ref{eq:eff-ham}-\ref{eq:fock}).}
    \label{fig:hea}
\end{figure}

It is noteworthy that each fragment 
state must be encoded with the highest possible accuracy so that the overall error of the full system can be minimized.
The accuracy of the individual fragment state (i.e. $|\psi^\mathrm{H_2}(\vec{\theta}) \rangle$, $|\psi^\mathrm{C_2H_2}(\vec{\theta}) \rangle$ etc) is measured by calculating  
the error in their energy with respect to the exact energy obtained via exact diagonalization of the fragment hamiltonians (Eq. \ref{eq:eff-ham}).
In Fig. \ref{fig:hea}(a), we have shown this error with an increasing number of layers in the HEA.
The error reaches $\sim$ 10$^{-8} \ \mathrm{E_h}$ 
with 6 layers for all molecular systems. 

The H$_4$ system serves as an intriguing model where the level of degeneracy can be smoothly tuned, ranging from a completely nondegenerate configuration to one with strong degeneracy.
At square geometry, the H$_4$ system shows a strong quasidegeneracy between the configurations $1a_1^21b_2^2$ and $1a_1^22a_1^2$.
This complicates the selection of individual H$_2$ subsystems, thereby necessitating a few additional layers to accurately encode the fragment states.
On the other hand, at the rectangular geometry where $r_1 =1.0$\AA \ and $r_2 =2.0$\AA, the quasidegeneracy fades away and the lowest energy configuration becomes $1a_1^21b_2^2$.

The CBD molecule in $D_{4h}$ symmetry has an open-shell singlet ground state, which undergoes pseudo-Jahn-Teller distortion to attain a $D_{2h}$ symmetry with a closed-shell singlet ground state. In $D_{2h}$ symmetry, the ground state of the molecule has a weak multiconfigurational character. In contrast, the ground state in $D_{4h}$ symmetry is a diradical, characterized by two degenerate singly occupied frontier MOs, thereby necessitating a multireference approach.
Fig. \ref{fig:hea}(a) shows the accuracy for preparing the individual C$_2$H$_2$ fragment states. 
For both $D_{2h}$ and $D_{4h}$ symmetries, the fragment HEA states can be prepared with an error of $\approx 10^{-8} \ \mathrm{E_h}$ using 6–8 layers. Moreover, the performance of the other two-qubit entangling blocks, such as full, circular, and pairwise, is also assessed for the square H$_4$ (see Fig. \ref{fig:hea}(b)), and we observe that most of the trends are quite similar.
The fidelity between the prepared and exact fragment states also indicates better state encoding with an increasing number of layers (see Supplementary Table 1).


\subsection{Comparison of Initial State Preparation}

\begin{figure}[!htb]
    \centering
    \includegraphics[width=\linewidth]{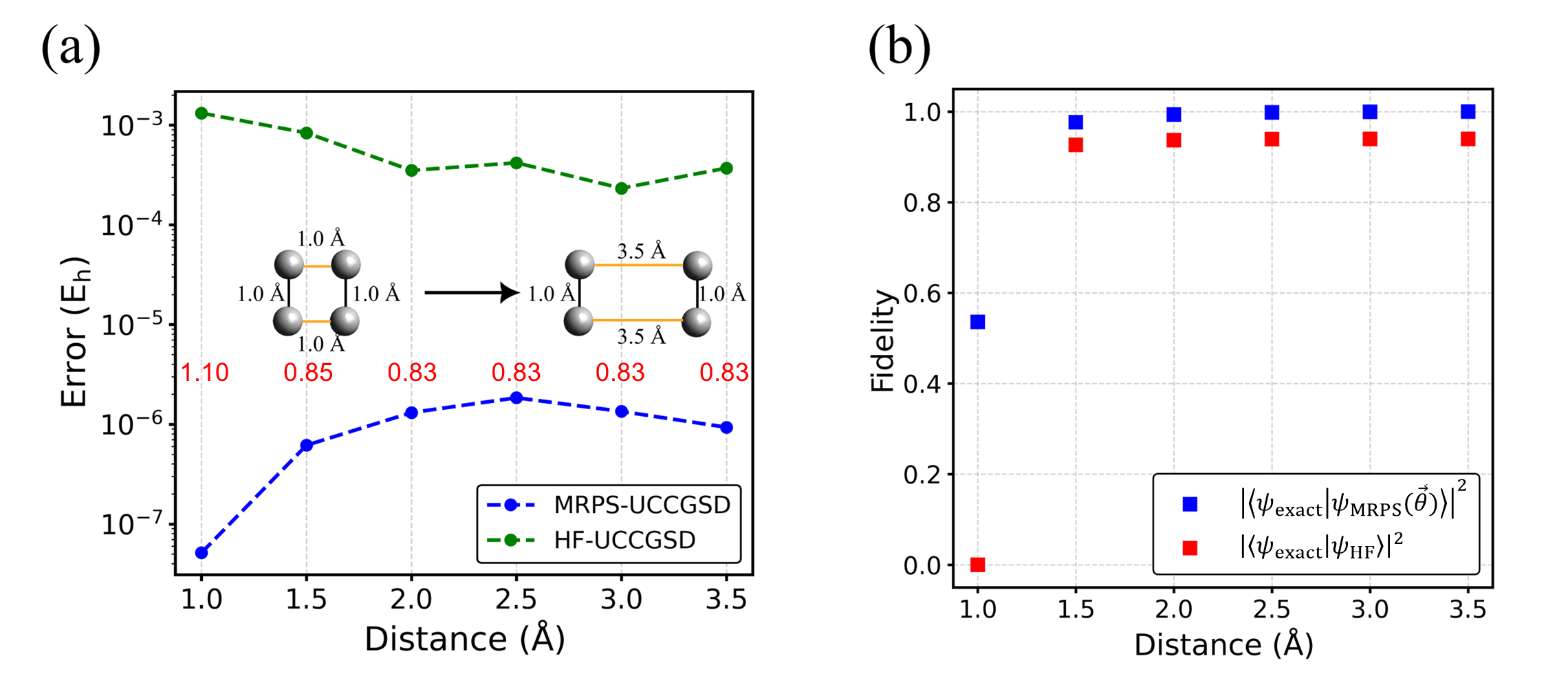}
    \caption{(a) Error (in Hartree) in the potential energy curve for the H$_4$ system calculated by full UCCGSD optimization starting with HF and MRPS initial states. The Shannon entropy values, indicating the degree of multiconfigurational character, are mentioned in red across these geometries.
 (b) The fidelities $|\langle \psi_\mathrm{exact} |\psi_\mathrm{MRPS}(\vec{\theta}) \rangle |^2$ and $|\langle \psi_\mathrm{exact} |\psi_\mathrm{HF} \rangle |^2$ shown along the potential energy curve for the H$_4$.}
    \label{fig:uccgsd-vqe}
\end{figure}

The need for initial state preparation via the multireference product state (MRPS) is evaluated by comparing UCCGSD results for the full system using two different reference states: HF and MRPS. The approach employing the HF reference is referred to as HF-UCCGSD, while that using the MRPS reference is denoted as MRPS-UCCGSD.
The error in the potential energy curve (PEC) for  H$_4$ (we stretch the $r_2$ distance, shown in Fig. \ref{fig:molecule}, from a square to rectangular geometry) is shown for HF-UCCGSD and MRPS-UCCGSD as compared to the exact energy in Fig. \ref{fig:uccgsd-vqe}a. 
For all points along the PEC, we run 10 independent simulations using randomly initialized UCCGSD parameters, and the mean values are shown.
We notice that the errors along the bond breaking coordinate are within $\approx 10^{-3}-10^{-4} \ \mathrm{E_h}$ for the HF reference. In comparison, it is well-converged to $\approx 10^{-6}-10^{-7} \ \mathrm{E_h}$ for the MRPS reference state. In other words, the accuracy increases from mHartree to $\mu$Hartree and beyond upon 
incorporating a multiconfigurational reference state. This is further noticed from the non-parallelity errors of 1.09 $\mathrm{mE_h}$ and $1.8 \ \mathrm{\mu E_h}$ for HF and MRPS references, respectively.
The extent of the multiconfigurational nature at each geometry is estimated from the Shannon entropy ($S=-\sum_i n_i \ ln(n_i)$, where $n_i$ is the occupation number of the natural orbital $i$) calculated on the exact wavefunction.  The Shannon entropy values for each geometry are shown in red in Fig. \ref{fig:uccgsd-vqe}a. This clearly shows that 
the entropy is highest at the square geometry and falls off after that. At this geometry, the MRPS reference results in the lowest errors as opposed to the HF reference wavefunction. This further demonstrates that the MRPS reference state effectively captures the multiconfigurational character of the system.

Almost all randomly initialized MRPS-UCCGSD simulations converge to an error of approximately 10$^{-7} \ \mathrm{E_h}$, whereas the HF-UCCGSD simulations—whether initialized with random, zero, or MP2 parameters—tend to plateau at errors in the range of 10$^{-3}$–10$^{-4} \ \mathrm{E_h}$ (see supplementary Fig. S1).
It is also worth noting that when the full-system circuit is constructed using only the HEA, the lowest error in energy is $\sim10^{-2} \ \mathrm{E_h}$ among 10 independent runs, since the full system HEA circuit is trapped in barren plateaus. This full system HEA circuit is simulated using 8 layers and linear entanglement blocks.

The deviation between MRPS-UCCGSD and HF-UCCGSD is further examined by evaluating the quality of the MRPS and HF wavefunctions relative to the exact wavefunction of the full system.
The fidelities $|\langle \psi_\mathrm{exact} |\psi_\mathrm{MRPS}(\vec{\theta}) \rangle |^2$ and $|\langle \psi_\mathrm{exact} |\psi_\mathrm{HF} \rangle |^2$ shown in Fig. \ref{fig:uccgsd-vqe}b demonstrate that the MRPS reference states possess multiconfigurational character and are therefore capable of accurately reproducing the exact wavefunction of the full system. This is more evident in the square H$_4$ system, where the HF wavefunction exhibits nearly zero overlap with the exact wavefunction, whereas the MRPS wavefunction achieves an overlap exceeding 0.5 with the exact wavefunction.

\subsection{Adaptive Methods for Inter-Subsystem Correlation Capture}

The total CNOT gate counts for UCCGSD on the HF and MRPS reference states are 2672 and 2596, respectively. This is computationally challenging for NISQ devices.
To reduce the CNOT gate counts, we employ ADAPT-VQE instead of full UCCGSD optimization. The approach where the ansatz is built using ADAPT-VQE with fermionic excitations starting from the HF initial state is referred to as HF-fermionic-ADAPT-VQE, whereas starting from the MRPS initial state 
is termed MRPS-fermionic-ADAPT-VQE.
We used an energy gradient threshold of 10$^{-8} \ \mathrm{E_h}$, which means the ADAPT iteration will continue until all the gradients are below this value.
\begin{figure}[!htb]
    \centering    \includegraphics[width=0.6\linewidth]{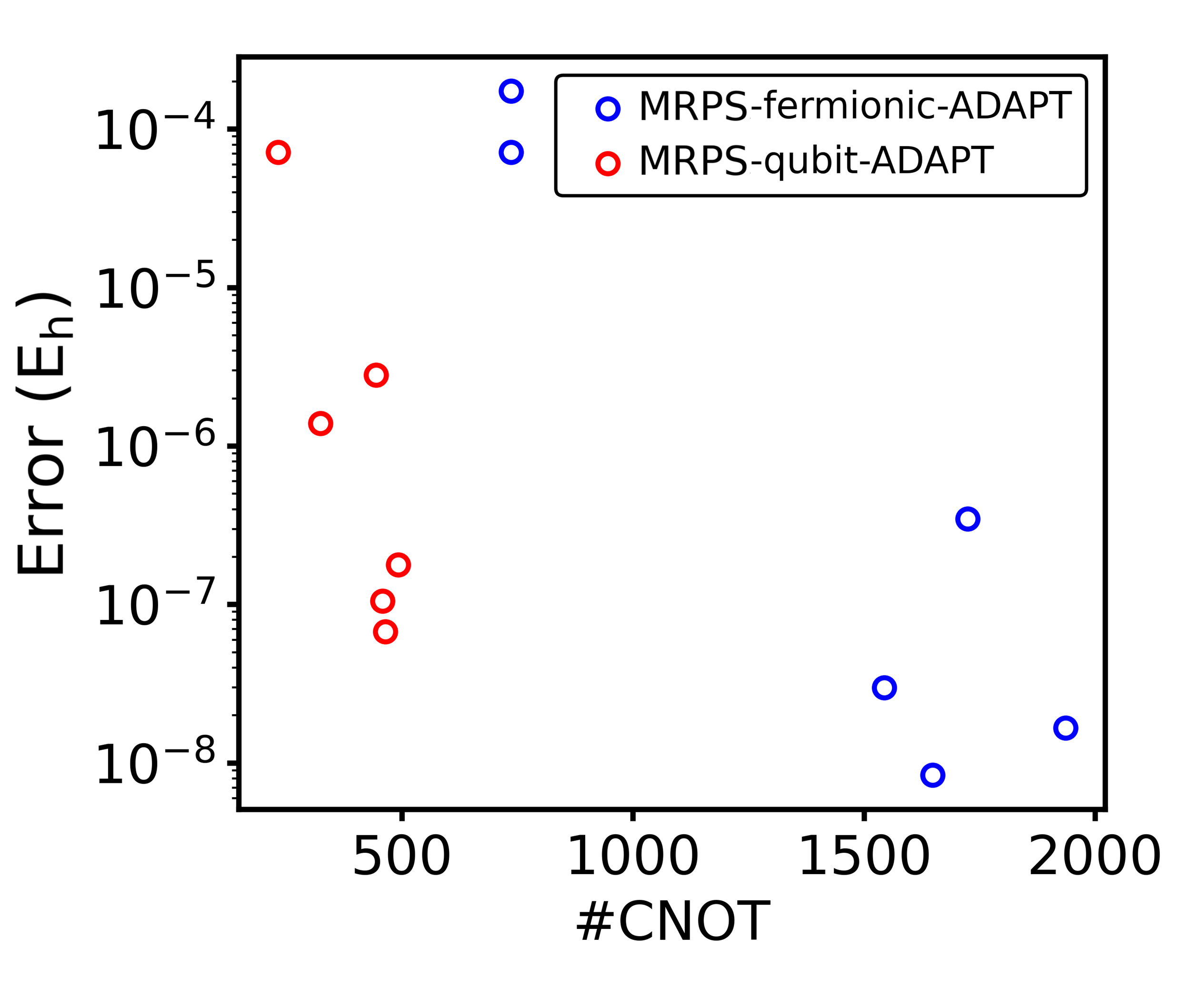}
    \caption{Error in energy (in Hartree) vs. CNOT gate counts for calculating the potential energy curve (PEC) of H$_4$ using MRPS-fermionic-ADAPT-VQE and MRPS-qubit-ADAPT-VQE. Different points refer to different geometries in the PEC.}
    \label{fig:fermi-qubit}
\end{figure}

The MRPS-fermionic-ADAPT-VQE demonstrates higher accuracy compared to HF-fermionic-ADAPT-VQE, specifically in the strongly correlated region (shown in supplementary Fig. S2). 
Furthermore, replacing the fermionic operator pool with the qubit pool using the MRPS initial state in ADAPT-VQE leads to a substantial reduction in the CNOT gate count (see Fig. \ref{fig:fermi-qubit}). 
The points in the figure correspond to various geometries along the PEC, and to achieve comparable accuracy in energy, MRPS-qubit-ADAPT-VQE consistently requires fewer CNOT gates across all these geometries.
This highlights the advantage of qubit-ADAPT over fermionic-ADAPT when starting from an MRPS reference state, consistent with the earlier observations made for the HF reference state.\cite{tang2021qubit}

%
%
%

\subsection{Potential Energy Profile of Strongly Correlated Molecules}
\begin{figure}[!htb]
    \centering    \includegraphics[width=0.6\linewidth]{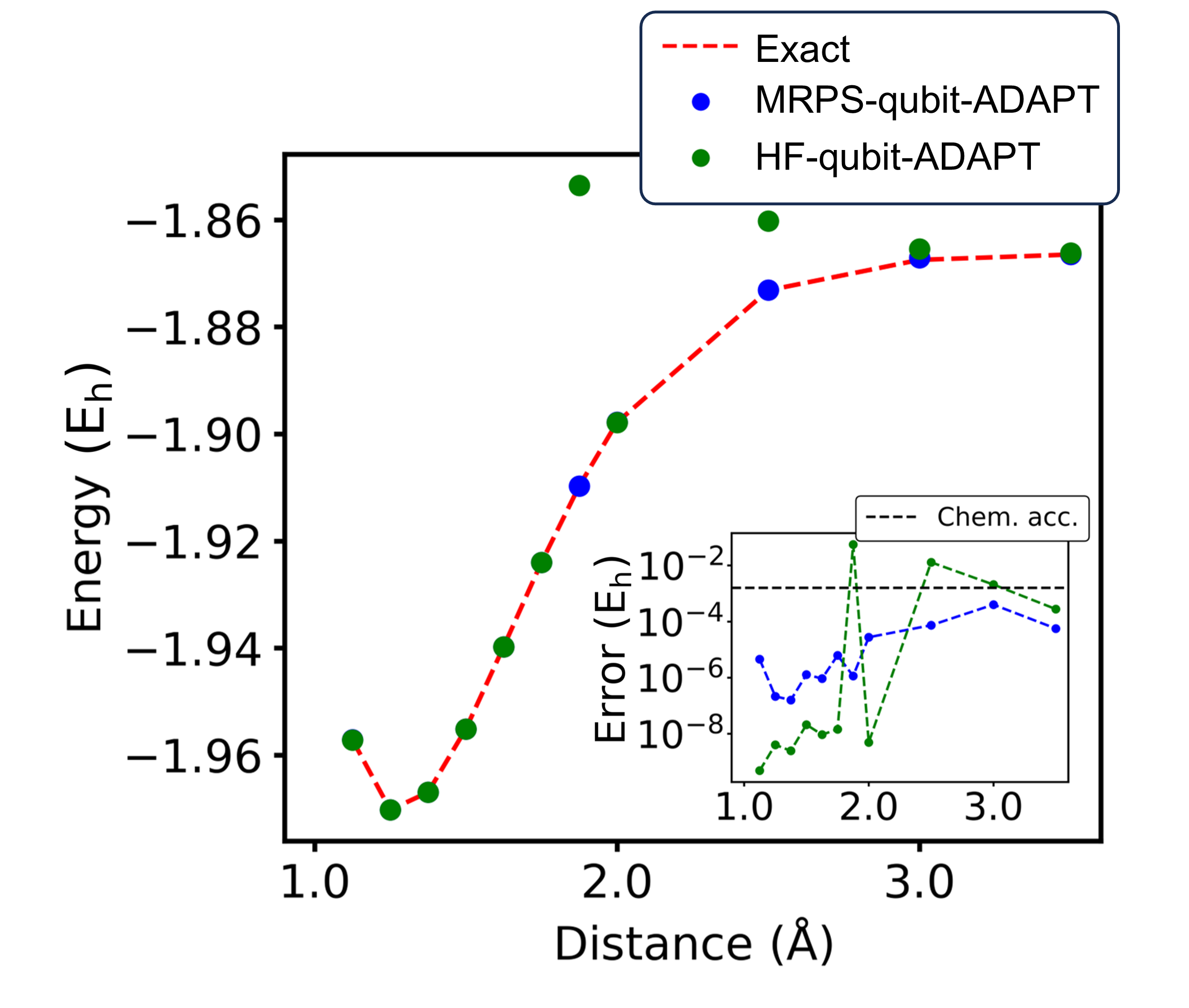}
    \caption{Potential energy curve for symmetric stretching of all four H--H bonds in H$_4$ square calculated with MRPS-qubit-ADAPT-VQE and HF-qubit-ADAPT-VQE. The inset shows the corresponding energy errors. The dashed black line in the inset refers to the chemical accuracy.}
    \label{fig:symm}
\end{figure}

Using MRPS-qubit-ADAPT-VQE, the PEC for symmetric stretching of all four H--H bonds in H$_4$ square is calculated (see Fig. \ref{fig:symm}). The corresponding errors are shown in the inset. 
The method demonstrated high accuracy, with energy errors consistently within the chemical accuracy across all points. 
Conversely, when using a HF initial state (i.e., the HF-qubit-ADAPT-VQE), the performance drops in the strongly correlated region.
Moreover, we achieve $\sim$80\% reduction in the CNOT gate counts by using qubit-ADAPT-VQE compared to full UCCGSD optimization.

\begin{figure}[!htb]
    \centering
    \includegraphics[width=0.9\linewidth]{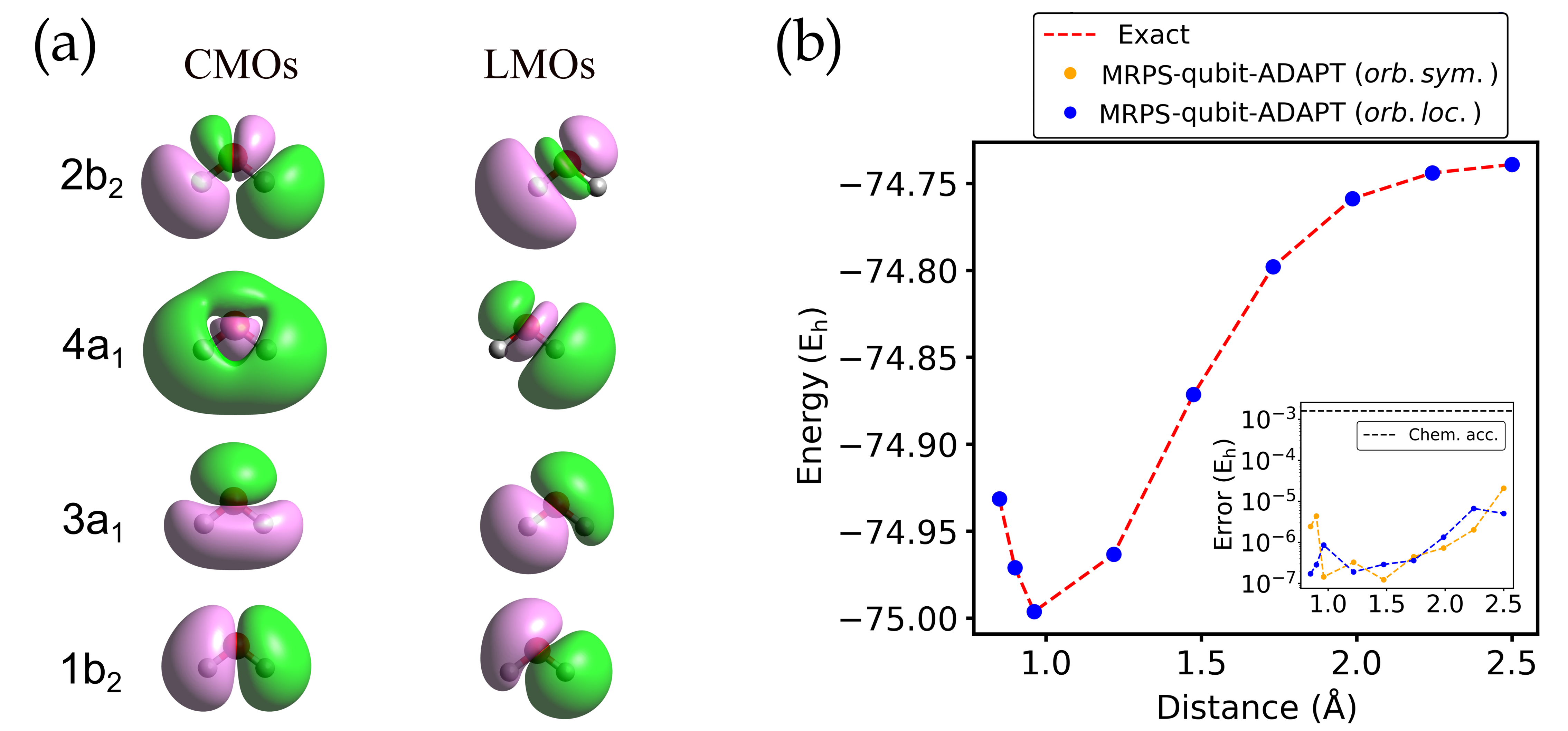}
    \caption{Bond breaking of water. (a) Canonical HF orbitals with the orbital symmetries considered in the active space for water (left), and the corresponding localized molecular orbitals (right). (b) Potential energy curve for the double dissociation of the O--H bonds in water calculated with MRPS-qubit-ADAPT-VQE, using orbital symmetries (\textit{orb. sym.}) and orbital localization (\textit{orb. loc.}) for preparing the fragment states. The inset shows the corresponding energy errors. The dashed black line in the inset refers to the chemical accuracy.}
    \label{fig:water}
\end{figure}

The performance of MRPS-qubit-ADAPT-VQE is also tested for the double dissociation of the O--H bonds in water molecule (Fig. \ref{fig:water}). The 
(4o,4e) active space comprising the CMOs is shown.
To prepare the fragment states, we used two distinct strategies. 
In the first, fragment states were constructed within the same orbital symmetries using a (2o,2e) active space—specifically within the (1b$_2$, 2b$_2$) and (3a$_1$, 4a$_1$) orbitals. 
The second approach involved localizing the CMOs, followed by preparing the fragment states in these localized orbital spaces (see LMOs in Fig. \ref{fig:water}a). For both strategies, the MRPS-qubit-ADAPT-VQE accurately reproduces the PEC, as shown in Fig. \ref{fig:water}b, achieving errors as low as 10$^{-7} \ \mathrm{E_h}$. Non-parallelity errors of $2.1\times10^{-2} \ \mathrm{mE_h}$ and $6.5\times10^{-3} \ \mathrm{mE_h}$ are observed, which shows that the bond dissociation energies can be estimated with errors of $\sim 10^{-3} \ \mathrm{mE_h}$ or $10^{-4}$ kcal/mol.
The CNOT gate requirement in these two strategies is compared in the supplementary Fig. S3.

\begin{figure}[!htb]
    \centering
    \includegraphics[width=.9\linewidth]{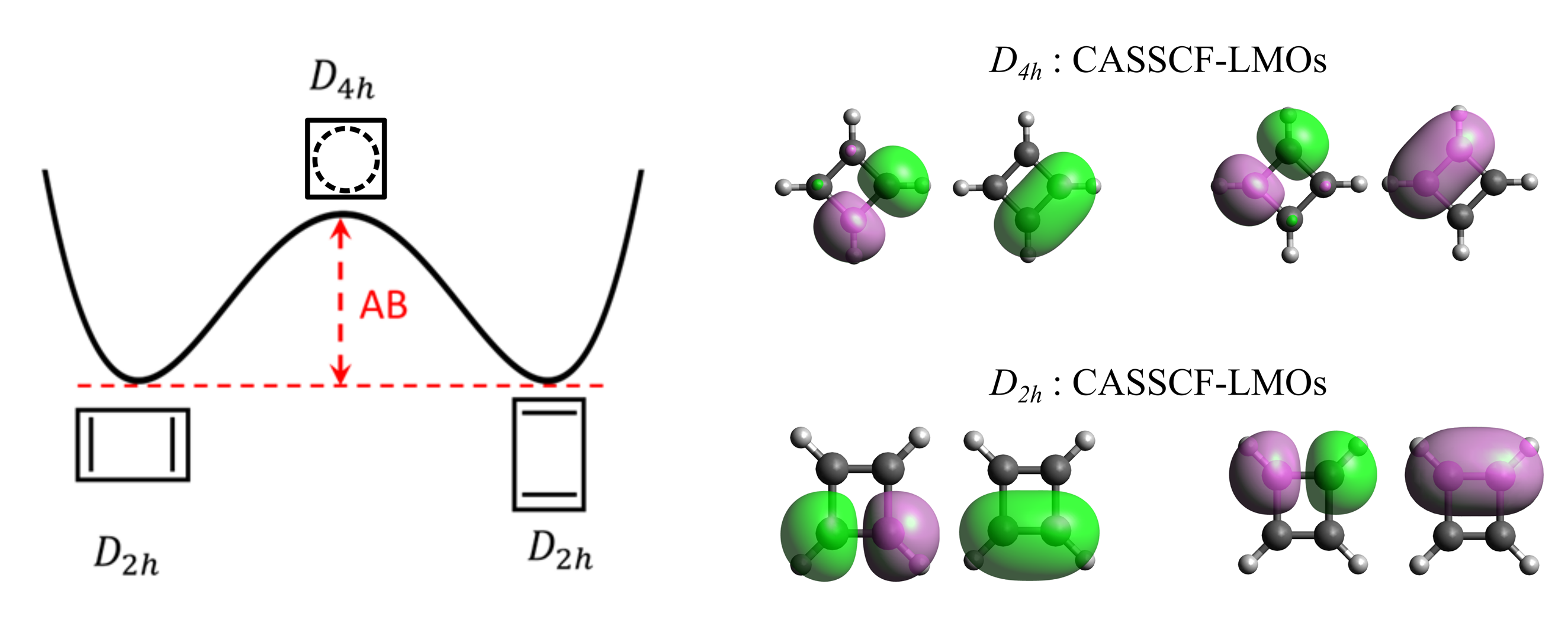}
    \caption{The graphical representation of the cyclobutadiene automerization reaction. The automerization barrier (AB) is depicted using a red arrow. The LMOs obtained by localizing the CASSCF orbitals for $D_{2h}$ and $D_{4h}$ geometries are also shown.}
    \label{fig:CBD_LMOs}
\end{figure}

We calculated the barrier of the automerization between two $D_{2h}$ rectangular CBDs (Fig. \ref{fig:CBD_LMOs}). The automerization barrier is defined as the difference between the ground state energies of $D_{2h}$ and $D_{4h}$ geometries.
Here, (4o,4e) active space in 6-31G basis consisting of the $\pi$ and $\pi ^*$ MOs is used. The fragment states are prepared on (2o,2e) space using orbital localization.
The ground states are calculated using MRPS-qubit-ADAPT-VQE. 
Table \ref{tab:cbd} shows the energies and automerization barriers obtained using the LMOs based on HF as well as CASSCF orbitals.
The MRPS-qubit-ADAPT-VQE method reproduces the energies with $\approx 10^{-7} \ \mathrm{E_h}$ error compared to the exact values.
However, using HF orbitals to obtain the LMOs which are used to compute the one- and two-electron integrals (in Eq. \ref{eq:eff-ham} and \ref{eq:fock}), MRPS-qubit-ADAPT-VQE gives automerization barrier of 23.7 kcal/mol as compared to the experimental value of 1.6$-$10 kcal/mol\cite{whitman1982limits}.
Therefore, to enable a meaningful comparison with the experimental data, it is necessary to use better orbitals for evaluating these integrals. Consequently, we employed the CASSCF orbitals for our calculations.
The LMOs based on the CASSCF orbitals are shown in Fig. \ref{fig:CBD_LMOs}.
With CASSCF orbitals, the MRPS-qubit-ADAPT-VQE estimates the barrier height to be 4.6 kcal/mol (see Table \ref{tab:cbd}), which is consistent with the experimental value.
The corresponding CNOT gate counts are provided in Table \ref{tab:cbd}, shown in parentheses.
For the $D_{4h}$ symmetry, the CNOT gate requirement is slightly higher when using the HF orbitals due to the strong multiconfigurational character of the wavefunction.
On the other hand, the automerization barrier is estimated to be around 56 kcal/mol using HF-qubit-ADAPT-VQE.
This is because the reference HF wavefunction fails to capture the strongly correlated open-shell ground state of the $D_{4h}$ geometry, and consequently overestimates the barrier height for the automerization reaction.

\begin{table}[H]
    \begin{center}        
    \begin{tabular}{p{2.2cm}|p{2.6cm} p{2.6cm}|p{2.6cm} p{2.6cm}} 
        \hline \hline 
        \textbf{} & \multicolumn{2}{|c|}{\textbf{HF orbitals}} & \multicolumn{2}{c}{\textbf{CASSCF orbitals}} \\[0.3cm]
         & \textbf{Exact} & \textbf{MRPS-qubit-ADAPT} & \textbf{Exact}  & \textbf{MRPS-qubit-ADAPT} \\[0.3cm]
        \hline \hline
        \textbf{$E(D_{2h})$} & -153.631656 & -153.631656 (224) & -153.650047 & -153.650047 (202) \\[0.3cm]
        \hline
        \textbf{$E(D_{4h})$} & -153.593841 & -153.593841 (416) & -153.642726 &  -153.642726 (154) \\[0.3cm]
        \hline
        \textbf{Barrier (kcal/ mol)} & 23.7 &  23.7 & 4.6  & 4.6 \\
        \hline \hline
    \end{tabular}
    \caption{The ground state energies (in Hartree) of $D_{2h}$ and $D_{4h}$ symmetry cyclobutadiene molecule calculated with the MRPS-qubit-ADAPT-VQE method. This is shown using HF as well as CASSCF orbitals. The corresponding automerization barrier is shown (in kcal/mol) in the last row. The number of CNOT gates required for each calculation is given in parentheses.}
    \label{tab:cbd}
    \end{center}
\end{table}

\section{Conclusions}



In this work, we have showcased several strategies for constructing multiconfigurational reference states, specifically designed to address the challenge of strong electron correlation in complex molecular systems. The cornerstone of our methodology is a bi-fold approach, which strategically partitions the task of modeling the electronic wavefunction into two distinct stages. This framework begins with a judicious problem decomposition, where the total system is divided into smaller, chemically intuitive fragments. For each of these fragments, we employ a hardware-aware ansatz, such as the Hardware Efficient Ansatz (HEA), to variationally prepare a highly entangled state. At this stage, the entanglement is present only within each individual fragment and not across them. The tensor product of these individual fragment states then forms a multireference product state (MRPS) that serves as the initial state. 

Upon this state, the second fold of our approach introduces the entanglement among the subsystems. This is achieved by applying a disentangled Unitary Coupled Cluster (dUCC) ansatz, using the MRPS as its reference initial state. The dUCC operators, which build the crucial inter-fragment entanglement, can either be chosen from the complete set of generalized single and double (GSD) excitations or selected adaptively to construct the most compact and efficient ansatz possible.
In summary, the bi-fold approach with suitable problem decomposition can effectively solve problems over exponentially large Hilbert spaces through several lower-dimensional sub-Hilbert spaces via disconnected 
optimization cycles, and the impact of this hierarchical strategy is evident in the accuracy of potential energy curves from mHartree to $\mu$Hartree levels. This, in turn, can be used to predict dissociation energies and reaction barriers that are well within the chemical accuracy. 

To improve the computational efficiency of these approaches, we have used ADAPT approaches with both fermionic and qubit excitations. We have noticed that qubit-ADAPT is more resource-efficient and amenable to NISQ devices. We have, therefore, demonstrated a versatile multireference approach that can be accurate and efficient on a quantum device and therefore, be applied to chemical problems of general interest.  The bi-fold scheme is expected to unlock the potential to design novel chemical space using near-term quantum devices. 

\begin{acknowledgement}
AC thanks INSPIRE and IACS for a senior research fellowship and a research associateship. SH acknowledges the Council of Scientific \& Industrial Research (CSIR) for their fellowship. DG thanks ANRF (Grant number: CRG/2023/001806) for generous funding. RM acknowledges the financial
support from the ANRF, Government of India (Grant Number: MTR/2023/001306).

\end{acknowledgement}

\begin{suppinfo}

\end{suppinfo}

\providecommand{\latin}[1]{#1}
\makeatletter
\providecommand{\doi}
  {\begingroup\let\do\@makeother\dospecials
  \catcode`\{=1 \catcode`\}=2 \doi@aux}
\providecommand{\doi@aux}[1]{\endgroup\texttt{#1}}
\makeatother
\providecommand*\mcitethebibliography{\thebibliography}
\csname @ifundefined\endcsname{endmcitethebibliography}  {\let\endmcitethebibliography\endthebibliography}{}

\end{document}